\documentclass{PoS}
\usepackage{amsmath,amssymb,array}

\title{
Top-Higgs Associated Production involving 
\boldmath{${A^0},\, {H^0}$} with Mass at 300 GeV
}

\ShortTitle{$t + S^0$-lite}

\author{\speaker{George W.-S. Hou}
   \thanks{Work done in collaboration with Masaya Kohda and Tanmoy Modak.}\\
         National Taiwan University\footnote{
             Home institute.
             }
        \\
        E-mail: \email{wshou@phys.ntu.edu.tw}}

\abstract{
We advocate two Higgs doublet model without $Z_2$ symmetry,
and focus on the extra top Yukawa couplings $\rho_{tt}$ and $\rho_{tc}$.
We show that $A^0$ and $H^0$ bosons are still allowed at 300 GeV mass,
where $cg \to tA^0$, $tH^0$ can lead to $tt\bar c$ same-sign top,
or top-assisted di-Higgs $th^0h^0$ signatures, resp.
As bonus material, we explore the constraint provided by current 4-top search,
but advocate direct search for triple-top generated by $cg \to tH^0/A^0 \to tt\bar t$
for $H^0,\, A^0$ above $t\bar t$ threshold,
and offer some insight on the mild but intriguing ``excess'' found by CMS
in $gg \to A^0 \to t\bar t$ at 400 GeV.
}

\FullConference{
European Physical Society Conference on High Energy Physics - EPS-HEP2019 -\\
			10-17 July, 2019\\
			Ghent, Belgium}

\begin{document}

\section{
Introduction: 2HDM without \boldmath{$Z_2$}}

In 1977, Glashow and Weinberg gave~\cite{Glashow:1976nt}  the ``edict'' of
Natural Flavor Conservation (NFC):
only one Yukawa matrix per mass matrix for two Higgs doublets.
This eliminated the worries of Flavor Changing Neutral Higgs (FCNH) couplings.
It arises automatically in supersymmetry,
where one doublet couples to $u$-type quarks, the other couples to $d$-type quarks,
which is two Higgs doublet model (2HDM), type II, 
usually implemented by a $Z_2$ symmetry.

Citing the Cheng-Sher pattern~\cite{Cheng:1987rs}
 of $\sqrt{m_i m_j}$ to control FCNH couplings, 
we pointed out~\cite{Hou:1991un} that $t \to ch^0$ or $h^0 \to t\bar c$, 
where $h^0$ is the lightest neutral scalar, are the best probes for such couplings.
The paper also stressed that the Cheng-Sher pattern reflects the emergent
fermion mass and mixing hierarchies, and called it Model III to
distinguish from usual 2HDM I \& II that invoke $Z_2$ symmetry. 
With observation of $h(125)$ in 2012, a second doublet is now highly plausible,
and it is imperative that we undertake experimental search.
Thinking back~\cite{Chen:2013qta} towards the 1991 proposal for $t \to ch$ search,
the ansatz of $\sqrt{m_i m_j}$ is seen as a ``scaffold'', and one should 
take the $2\times 2$ Yukawa matrix of the top-charm sector seriously.
For the four parameters, $\rho_{ct}$ is constrained by flavor physics 
to be rather small~\cite{Altunkaynak:2015twa}, and $\rho_{cc}$ should not be 
much larger than $\lambda_{cc} = \sqrt2 m_c/v < 0.01$, where $v = 246$ GeV. 
The two remaining extra Yukawa couplings $\rho_{tc}$ and $\rho_{tt}$ 
can be ${\cal O}(1)$, where $\rho_{tc}$ can induce $t \to ch^0$ decay, but 
modulated by the $h^0$--$H^0$ mixing angle $\cos(\beta - \alpha)$.
From here we see that FCNH couplings for $h^0$ can be 
suppressed by $h^0$--$H^0$ mixing.

With emergent ``alignment'', that $\cos(\beta - \alpha)$  (called $\cos\gamma$ from now on) 
appears to be rather small because $h^0$ is rather close to the SM Higgs boson,
we understand the non-observation so far of the $t \to ch^0$, $t \to uh^0$ 
and $h^0 \to \mu\tau$ decays are manifestations of alignment.
A discussion synthesizing mass-mixing hierarchy and approximate alignment
as replacement of NFC can be found in Ref.~\cite{Hou:2017hiw}.
We shall refer to 2HDM without $Z_2$, i.e. with extra Yukawas, as g2HDM.

In our previous EPS-HEP proceedings~\cite{Hou:2017kmo}, 
we emphasized that the extra Yukawa couplings 
$\rho_{tt}$ and $\rho_{tc}$, naturally ${\cal O}(1)$ and complex, 
could drive~\cite{Fuyuto:2017ewj} electroweak baryogenesis (EWBG).
The $\rho_{tt}$ coupling is the robust driver, but if it is less than 
a few \%, then an ${\cal O}(1)$ $\rho_{tc}$ 
with near maximal phase can be a backup option.
The present report arose from our invited talk~\cite{GHou2018} 
at Moriond QCD 2018, where we used a diagram from our 1997 
study of $cg \to tA^0$~\cite{Hou:1997pm} that gives same-sign top signature.
Expecting this lighter $A^0$ case to be ruled out by LHC data,
to our surprise, we found~\cite{Hou:2018zmg} the answer in the negative:
an $A^0$ as light as 300 GeV is still viable!
Replacing $A^0$ by $H^0$, the $CP$ even exotic scalar,
we explore the viable parameter space for 
top-assisted di-Higgs production~\cite{Hou:2019qqi} via $cg \to tH^0 \to th^0h^0$.
As bonus material, we further discuss~\cite{Hou:2019gpn} the 
4-top search constraint on triple-top, and mention 
an ``excess'' at 400 GeV~\cite{Sirunyan:2019wph} 
in scalar $t\bar t$ resonance search by CMS.

\section{
\boldmath{$cg \to tA^0 \to tt\bar c$}:  allowed for \boldmath{$m_{A^0} \sim$} 300 GeV}

Without a $Z_2$ symmetry, the coupling of $A^0$ to fermions is
 (no $\cos\gamma$ suppression)
\begin{equation}
\frac{i}{\sqrt{2}} \sum_{F = u, d, e}
 {\rm sgn}(Q_F) \rho^F_{ij} \bar F_{iL} F_{jR} A^0  +{\rm h.c.},
\end{equation}
where $i,j =1,2,3$ are generation indices, ${\rm sgn}(Q_F) = +1$ ($-1$) 
for $F = u$ ($F=d, e$), and $\rho^F$ are $3\times 3$ complex matrices.
In checking the original proposal for $cg \to tA^0$ production~\cite{Hou:1997pm},
we considered~\cite{Hou:2018zmg} a light $A^0$ in the range 
$200~{\rm GeV} < m_{A^0} < 340~{\rm GeV}$,
i.e. below $t\bar t$ threshold, allowing $A^0 \to t\bar c + \bar tc$ to be dominant.
However, to avoid the $A^0 \to Zh$ constraint from
ATLAS~\cite{Aaboud:2017cxo} and CMS~\cite{Sirunyan:2019xls},
we assume $|\rho_{tt}| \ll 1$ to suppress  $gg \to A^0$,
even though the $AZh$ gauge coupling is $\cos\gamma$, or alignment, suppressed.
This means our discussion can be extended somewhat to beyond the $t\bar t$ threshold.
Note that one still has the $\rho_{tc}$-driven EWBG~\cite{Fuyuto:2017ewj} 
even in this case. 
To simplify discussion, in the following numerics we set 
all $\rho^F_{ij} = 0$  except for $\rho_{tc}$, 
and assume the alignment limit where $c_\gamma = 0$,
so that $\mathcal{B}(A^0\to t \bar c) = 50\%$.
The $H^0$, $H^+$ bosons are treated\footnote{
 In this and the next section, we skip the discussion of near degenerate
 $A^0$ and $H^0$ bosons. See original papers.
}
 as somewhat heavier.

\begin{figure*}[t]
\center
\includegraphics[width=.45 \textwidth]{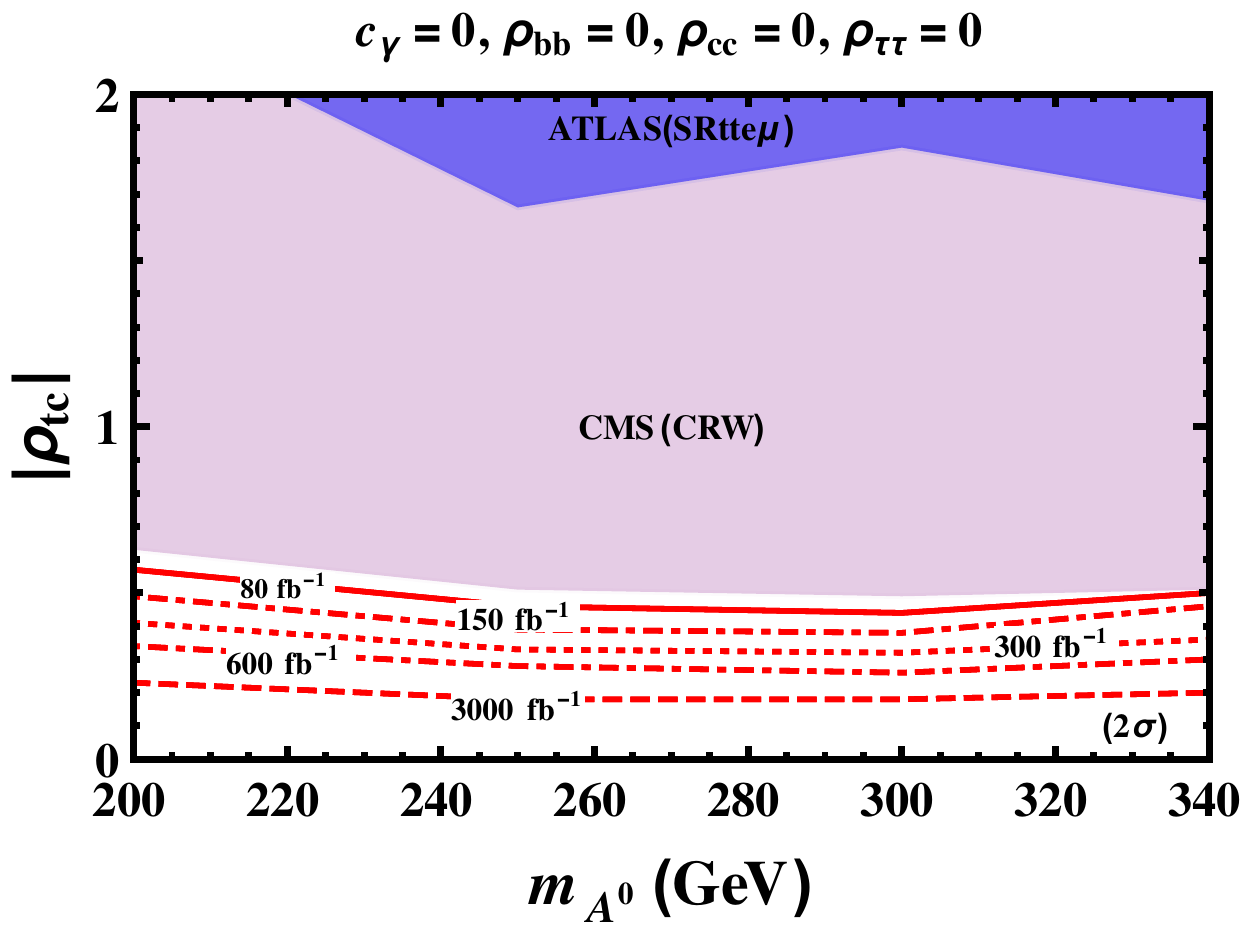}
\includegraphics[width=.45 \textwidth]{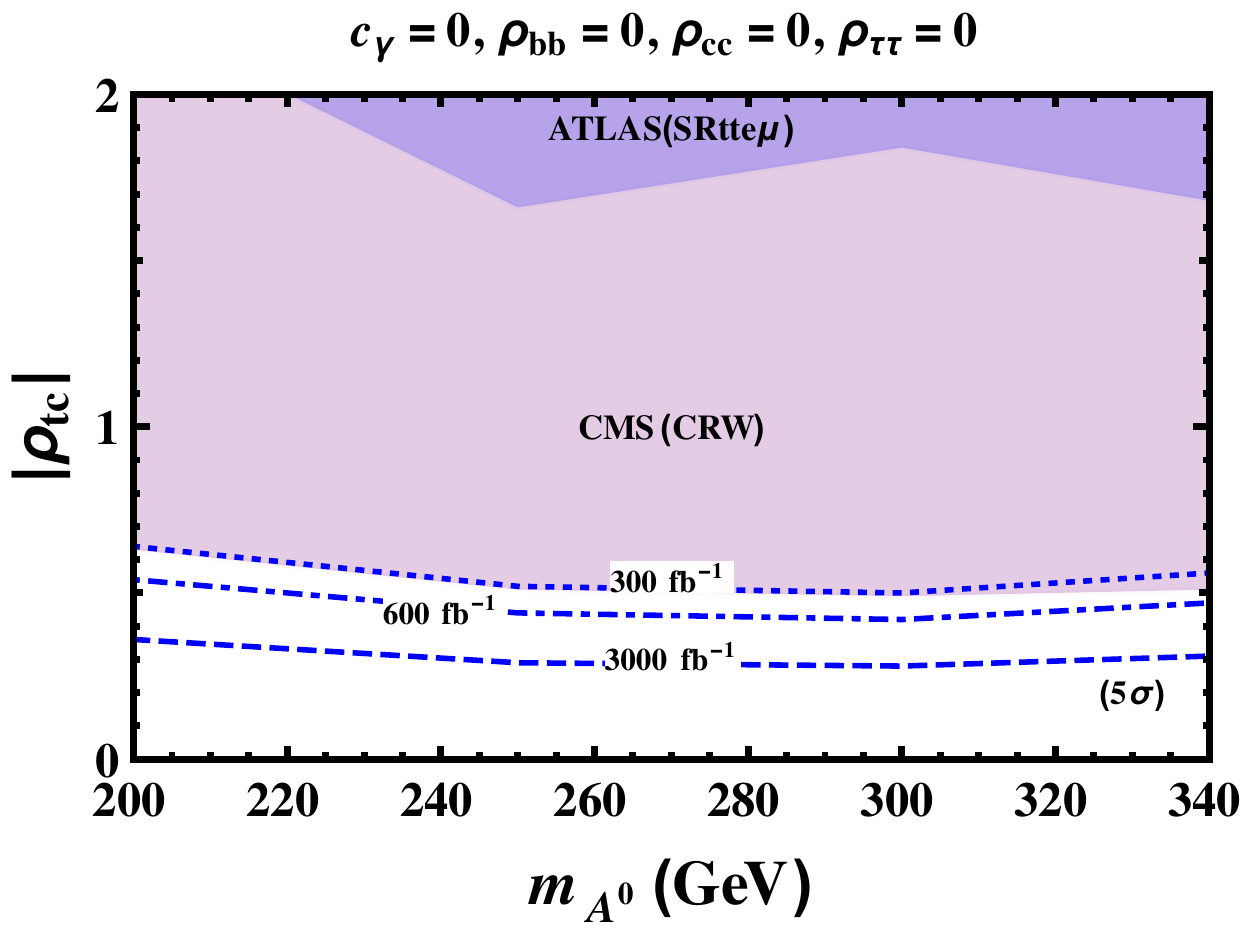}
\caption{
[left] 2$\sigma$ exclusion and [right] 5$\sigma$ discovery 
for $|\rho_{tc}|$ by SS$tt$ for various $\int {\cal L}dt$ at 14 TeV,
with SRtt$e\mu$ of ATLAS $cc \to tt$ search (blue/dark) and 
CRW of CMS $t\bar t t \bar t$ search (purple/light) overlaid.
}
\label{exclusion}
\end{figure*}

Assuming $\rho_{tc} = 1$, the cross section for $cg \to tA^0$ at 14 TeV is 
1 fb for $m_{{A^0}}$ at 200 GeV, and 1.3 fb for 340 GeV after selection cuts.
For the same-sign top plus jet (SS$tt$) signature, i.e. same-sign di-lepton plus jets,
the main background is $t\bar tW$ production.
ATLAS has a dedicated $qq\to tt$ search~\cite{Aad:2015gdg} 
in various signal regions (SRs), where we find SRtt$e\mu$,
or $e\mu$ final state from both tops decaying semileptonically,
giving the best limit on $\rho_{tc}$.
On the other hand, 
the most relevant constraint comes from the $t\bar t t\bar t$ search for CMS, which 
has been recently updated to 137 fb$^{-1}$~\cite{Sirunyan:2019wxt}, or full Run~2 data, 
from an earlier analysis~\cite{Sirunyan:2017roi} based on 2016 data. 
Based on the number of leptons, jets or $b$-tagged jets, 
CMS defines eight SRs plus two control regions (CRs) for background.
We find CRW for $t\bar tW$ background gives the best limit.
In the left panel of Fig.~\ref{exclusion} we give the 2$\sigma$ exclusion limits
for various integrated luminosities,
and the 5$\sigma$ discovery reaches are given in the right panel.
For comparison, we overlay the constraints from
 SRtt$e\mu$ of  ATLAS $cc \to tt$ search~\cite{Aad:2015gdg} from Run 1, and
 CRW ($t\bar tW$ control region)  of 4-top search by CMS~\cite{Sirunyan:2019wxt}
with full Run 2 data. 
Note that this plot is an update of our published work~\cite{Hou:2018zmg}
based on the earlier CMS PAS.

It should be clear that there is a lot of room for improving the SS$tt$ search,
but the constraint from full Run 2 4-top search is pretty good.
However, a dedicated search would be better.
It is also interesting to note that $cg \to tA^0 \to tt\bar c$ might be discovered
at 5$\sigma$ level for $\rho_{tc}$ around 0.6 with full Run 2 plus Run 3 data.
The HL-LHC could push down to smaller $\rho_{tc}$ values,
although it may become less pertinent for EWBG.
This highlights the value of a timely dedicated search.

We find it mildly surprising that a relatively light $A^0$ at or below $t\bar t$ threshold 
with FCNH coupling $\rho_{tc}$ is not yet ruled out by LHC data.
Direct search -- which probes EWBG! -- is encouraged.

\section{
\boldmath{$cg \to tH^0 \to th^0h^0$}:  allowed for \boldmath{$m_{H^0} \sim$} 300 GeV}

If in fact the $H^0$ is the lighter exotic scalar, one should consider
$cg \to tH^0$, again assuming $|\rho_{tt}| \ll 1$. 
Besides the discussion above, a new signature would be 
$H^0 \to h^0h^0$, leading to $cg \to tH \to thh$~\cite{Hou:2019qqi},
or top-assisted di-Higgs production.
Given the general interest in di-Higgs search at the LHC, 
this sounds interesting in its own right.
%

The $CP$-even scalars $h$, $H$ and $CP$-odd scalar $A$ couple to fermions 
by
\begin{equation}
 -\frac{1}{\sqrt{2}} \sum_{F = u, d, e}
 \bar F_{iL} \bigg[\big(-\lambda^F_{ij} s_\gamma + \rho^F_{ij} c_\gamma\big) h 
 +\big(\lambda^F_{ij} c_\gamma + \rho^F_{ij} s_\gamma\big)H -i ~{\rm sgn}(Q_F) \rho^F_{ij} A\bigg]  F_{jR}
   +{\rm h.c.},
\end{equation}
where $\lambda^F$ are the usual diagonal Yukawa matrices,
and we take $c_\gamma= \cos\gamma$, $s_\gamma= \sin\gamma$ as shorthand.
We see that in the alignment limit of $c_\gamma \to 0$,
couplings of $h$ become diagonal, i.e. approach SM Yukawa couplings, 
while $H$, $A$ couple via extra $\rho^F$ Yukawa matrices.
We again exploit the $\rho_{tc}$ coupling of $H$ for its associated production with top.

To discuss $Hhh$ coupling, we need to consider the Higgs potential,
which we assume is $CP$-conserving ($CP$ violation only through Yukawa).
It can be written in the Higgs basis as~\cite{Hou:2017hiw}
\begin{align}
  V(\Phi,\Phi')
   & = \mu_{11}^2|\Phi|^2 + \mu_{22}^2|\Phi'|^2 - (\mu_{12}^2\Phi^\dagger\Phi' + h.c.)
    + \frac{\eta_1}{2}|\Phi|^4 + \frac{\eta_2}{2}|\Phi'|^4 + \eta_3|\Phi|^2|\Phi'|^2  \nonumber \\
  & \quad + \eta_4 |\Phi^\dagger\Phi'|^2
  + \bigg[\frac{\eta_5}{2}(\Phi^\dagger\Phi')^2
     + \left(\eta_6 |\Phi|^2 + \eta_7|\Phi'|^2\right) \Phi^\dagger\Phi' + h.c.\bigg],
\label{pot}
\end{align}
where all parameters are real, 
$v$ arises from the doublet $\Phi$ via $\mu_{11}^2=-\frac{1}{2}\eta_1 v^2$, 
while $\left\langle \Phi'\right\rangle =0$, hence $\mu_{22}^2 > 0$.
The ``soft-breaking'' parameter is eliminated by a second minimization condition,
$\mu_{12}^2 = \frac{1}{2}\eta_6 v^2$, which  reduces 
the total number of parameters to nine~\cite{Hou:2017hiw},
just one more compared to 2HDM II.
For $c_\gamma$ small but not infinitesimal, 
one has  
\begin{align}
c_\gamma \simeq \frac{-|\eta_6| v^2}{m_H^2 -m_h^2},
\end{align}
where another relation connects the proximity of $m_h^2$ to $\eta_1 v^2$.
One finds approximate alignment i.e. small $c_\gamma$ 
can be attained~\cite{Hou:2017hiw} without requiring $\eta_6$, $\eta_1$ to be small,
which is an important check for the prerequisite of ${\cal O}(1)$ Higgs quartics
for sake of first order electroweak phase transition~\cite{Fuyuto:2017ewj}.

The $Hhh$ coupling is the coefficient of the
$\lambda_{Hhh}H h^2$ term derivable from Eq.~\eqref{pot}, 
\begin{align}
\lambda_{Hhh} & =  \frac{v}{2} \bigg[3 c_\gamma s_\gamma^2 \eta_1
   + c_\gamma (3c_\gamma^2 - 2) \eta_{345} + 3 s_\gamma( 1-3c_\gamma^2)\eta_6
   +3s_\gamma c_\gamma^2 \eta_7\bigg] \nonumber \\
 & \simeq  \frac{c_\gamma}{2}v \bigg[3\frac{m_H^2}{v^2} - 2\eta_{345}
                  + 3\mbox{sgn}(s_\gamma) c_\gamma \eta_7
                  + \mathcal{O}(c_\gamma^2)\bigg],
\label{lHhhapp}
\end{align}
where $\eta_{345}=\eta_3+\eta_4+\eta_5$,
and the second step follows for small $c_\gamma$,
which shows that $\lambda_{Hhh} \to 0$ as $c_\gamma\to 0$.
To enhance $cg\to tH\to t hh$, sizable $\lambda_{Hhh}$ is needed,
and $\eta_{345} <0$ may be preferred so the first two terms add up.
The Higgs quartic parameters $\eta_1$ and $\eta_{3-6}$ can be 
expressed~\cite{Hou:2017hiw,Hou:2019qqi} in terms of 
$m_h$, $m_A$, $m_H$, $m_{H^+}$, $\mu_{22}$,
all normalized to $v$, as well as the mixing angle $\gamma$,
but $\eta_2$ and $\eta_7$ do not enter scalar masses.
We do not go into the details~\cite{Hou:2019qqi} here, but with 
$v = 246$ GeV and $m_h = 125$~GeV fixed, 
in the following numerics, we scan over
$\mu_{22} \in [0, 700]$  GeV, 
$\eta_2 \in [0, 3]$, $ \eta_7 \in [-3, 3]$, 
$m_H \in [250, 500]$~GeV, 
$m_{H^+} \in [300, 600]$ GeV,
$m_{A}\in [250, 500]$ GeV  with $m_H < m_A,\,m_{H^+}$,
and $\gamma$ values that satisfy $c_\gamma \in [0, 0.2]$. 
We identify $\eta_{1-7}$ with $\Lambda_{1-7}$ of 2HDMC~\cite{Eriksson:2009ws},
which we use to enforce perturbativity (conservatively impose $|\eta_i| < 3$), 
tree unitarity and positivity.

\begin{figure*}[t!]
\center
\includegraphics[width=.42 \textwidth]{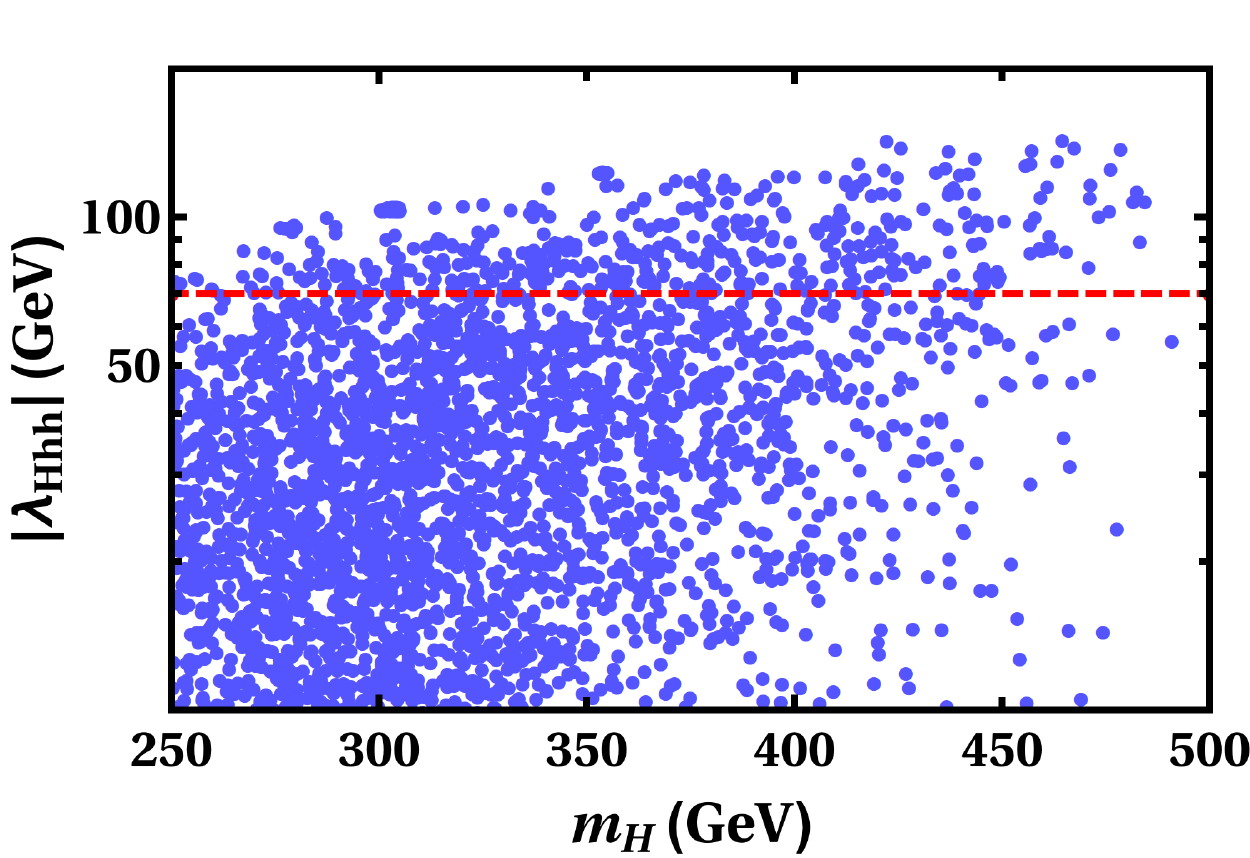}
\includegraphics[width=.408 \textwidth]{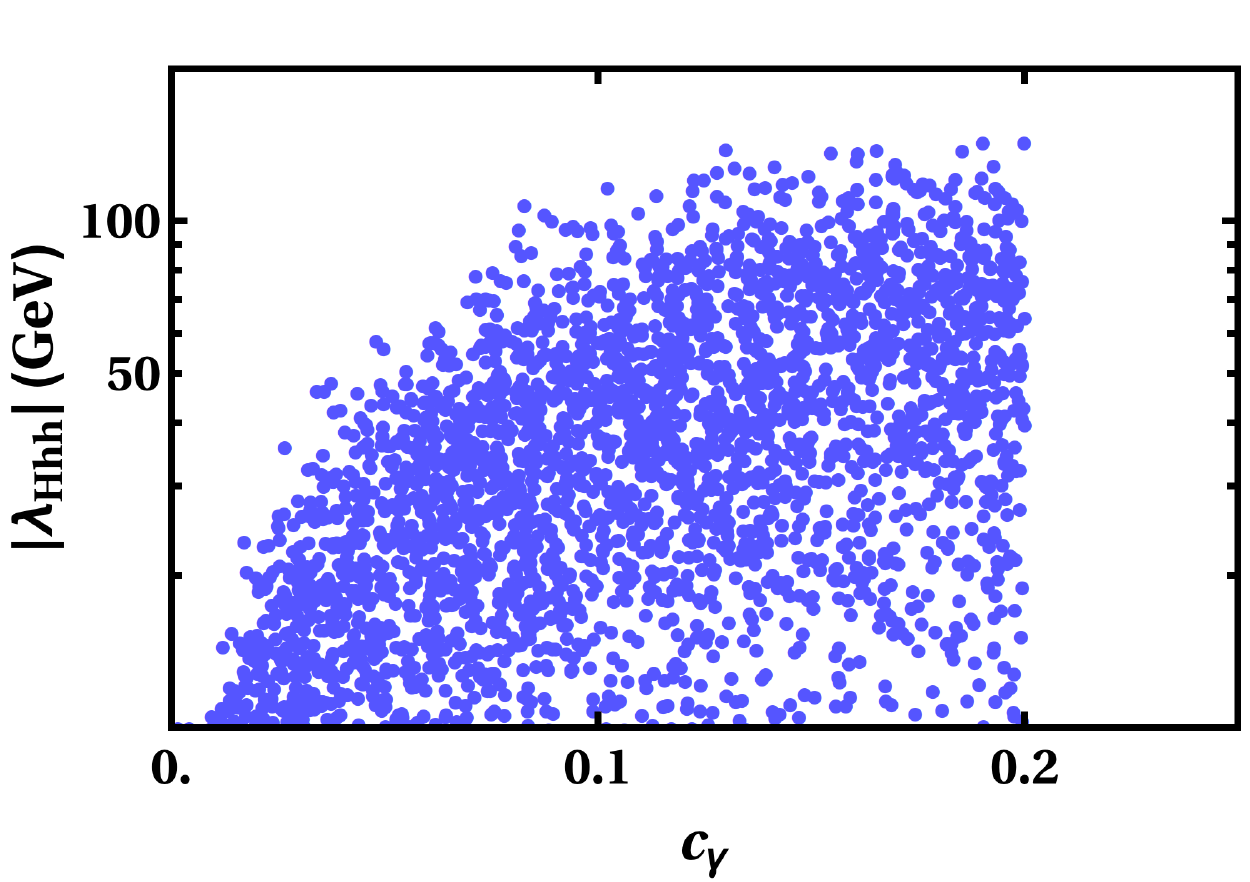}
\caption{
$\lambda_{Hhh}$ vs $m_H$ and $c_\gamma$ plots for 
the scan points that pass perturbativity, tree-level unitarity 
and positivity through 2HDMC, where $|\eta_i| < 3$ is maintained.
The $T$ parameter constraint is also imposed.
} 
\label{lH0hh}
\end{figure*}

\begin{table}[t]
\centering
\vskip0.3cm
{\scriptsize
\begin{tabular}{l |c| c| c| c | c | c| c | c |c| c|c| c| c}
\hline
  & $\eta_1$ &  $\eta_2$   &  $\eta_3$   & $\eta_4$  & $\eta_5$ &$\eta_{345}$ & $\eta_6$  & $\eta_7$  & $m_{H^\pm}$  & $m_A$ & $m_H$ 
& $|\lambda_{Hhh}|$ & $\frac{\mu_{22}^2}{v^2}$\\
 &&&&&&&&& (GeV) & (GeV) & (GeV) & (GeV)&\\ 
\hline
\hline
1         & 0.287& 3.00 & $-0.188$ & 2.04& $-2.56$ & $-0.704$ & $-0.172$ & 0.557& 303& 481& 280 & 97 & 1.61  \\
2         & 0.294 & 2.78 &  0.269& 2.10& $-2.95$ & $-0.581$ &  $-0.21$ & 0.633& 340& 518& 304 & 104 & 1.77  \\
3         & 0.309& 2.98 & $-0.017$ & 2.42& $-2.73$ & $-0.328$ & $-0.301$ & 0.881& 363& 536& 354 & 123& 2.18  \\
&&&&&&&&&&&&&\\
\hline
\hline
\end{tabular}
}
\caption{Parameter values for three benchmark points.}
\label{bench}
\end{table}

The final ``scan points'' within $2\sigma$ error of $T$ parameter constraint
are plotted in Fig.~\ref{lH0hh}, where the horizontal dashed line drawn at
$|\lambda_{Hhh}| \sim 70$ GeV is to illustrate that $\lambda_{Hhh}$ 
can be sizable over a finite parameter region. 
Note that the condition $m_H < m_A,\,m_{H^+}$ 
forbids the decays $H \to AZ,\, H^\pm W^\mp$.
From the scan points, we select three favorable benchmark points\footnote{
A second set of benchmarks as well as more discussions are
 given in Ref.~\cite{Hou:2019qqi}.
}
for $m_H \lesssim 350$ GeV,
with $\lambda_{Hhh}$ around or slightly above 100 GeV, as given in Table 1.
All three benchmark points corresponds to $c_\gamma = 0.169$
($s_\gamma = -0.986$), which is possible from Fig.~1[right].
They also correspond to $\rho_{tc} = 0.54$,
with ${\cal B}(H \to hh)$ between 23\% to 24\%.
Note that the $\rho_{tc}$ and $c_\gamma$ values
are consistent with $t \to ch$ bounds.
The $H \to t\bar c + \bar t c$ branching fraction is around 70\%.

\begin{table}[b]
\centering
\scriptsize
\begin{tabular}{c |c| c| c| c | c| c| c}
\hline
&&&&&&&\\ 
               & $t\bar t$   & Single-$t$    &  $t\bar t h$ & $4t$   & $t\bar t W$ &  $t\bar t Z$ & Others             \\
                  &   (fb)      &(fb)     &   (fb)       &   (fb) &   (fb)  & (fb)   & (fb)  \\      
\hline
\hline
  1 & 6.70 & 1.01 & 1.01 & 0.016 & 0.022 & 0.234 & 0.007 \\  
  2 & 7.42 & 1.01 & 1.12 & 0.019 & 0.022 & 0.262 & 0.008 \\  
  3 & 7.94 & 1.52 & 1.14 & 0.024 & 0.020  & 0.268 & 0.008 \\  
&&&&&&&\\
\hline
\hline
\end{tabular}
\caption{Background cross sections after selection cuts at $\sqrt{s}=14$ TeV.}
\label{bkgcomp}
\end{table}

With these three benchmark points, 
we study the signature of $cg \to tH \to b\ell\nu + 4b$~\cite{Hou:2019qqi}.
Imposing lepton, missing $E_T$ cuts and requiring at least 5 jets
 (with at least 4 $b$-tagged), the main background cross sections
are given in Table~\ref{bkgcomp}.
For benchmark 1, 2, 3, we find 
signal cross sections of 0.396, 0.38, 0.288 fb, 
total background cross sections of 9.00, 9.86, 10.92 fb, and 
 significance of 3.2, 2.9, 2.1 (7.2, 6.6, 4.8) for
600 (3000) fb$^{-1}$, respectively.
We see that, though $H \to hh$ needs finite $h$-$H$ mixing 
as well as ${\cal O}(1)$ extra Higgs quartics, as it now stands,
the $thh$ or top-assisted di-Higgs signature has 
non-negligible discovery potential at the LHC.

\section{4t on Triple-Top; and ``Excess'' in $A^0 \to t\bar t$ at 400 GeV?}

In this section we present some ``bonus'' material from our more recent
work~\cite{Hou:2019gpn} on 4-top constraints on g2HDM, 
and insight on some $t\bar t$ news.

One peculiarity of Nature is that, while SM
single-top and $t\bar t$ enjoy the sizable cross sections at 14 TeV
of roughly 0.2 and 0.6 nb, respectively,
triple-top cross section at roughly 2 fb~\cite{Barger:2010uw}
 is even smaller than 4-top production at just above 10 fb. 
As such, the ATLAS and CMS experiments have been pursuing 4-top search, 
where CMS has recently updated their previous result based on 
36 fb$^{-1}$~\cite{Sirunyan:2017roi} to full Run~2 data of 
137 fb$^{-1}$~\cite{Sirunyan:2019wxt},
clearly zooming in on 4-top cross section measurement
(epitomized by the measured central value agreeing with theory projection).
However, the search for triple-top has been bypassed.

We have focused so far on a lighter $A$ or $H$, i.e. below $t\bar t$ threshold,
which are actually spin-offs from our earlier triple-top work.
For $H$ and $A$ above $t\bar t$ threshold, we proposed~\cite{Kohda:2017fkn} 
two years ago the process $cg \to tH,\,tA \to tt\bar c,\, tt\bar t$,
where the same-sign top signature arises from decay to $t\bar c$ via $\rho_{tc}$,
while the triple-top signature arises from $H,\,A \to t\bar t$ decay via $\rho_{tt}$.
Assuming $\rho_{tc}$ (needed for production) and $\rho_{tt}$ are ${\cal O}(1)$,
the triple-top cross section can be larger than 4-top, extending even to pb.
In the classic argument that a suppressed SM effect makes the
New Physics search more compelling, we have urged all along
a targeted, direct search for triple-top at the LHC.

\begin{figure*}[b]
\center
\includegraphics[width=0.36 \textwidth]{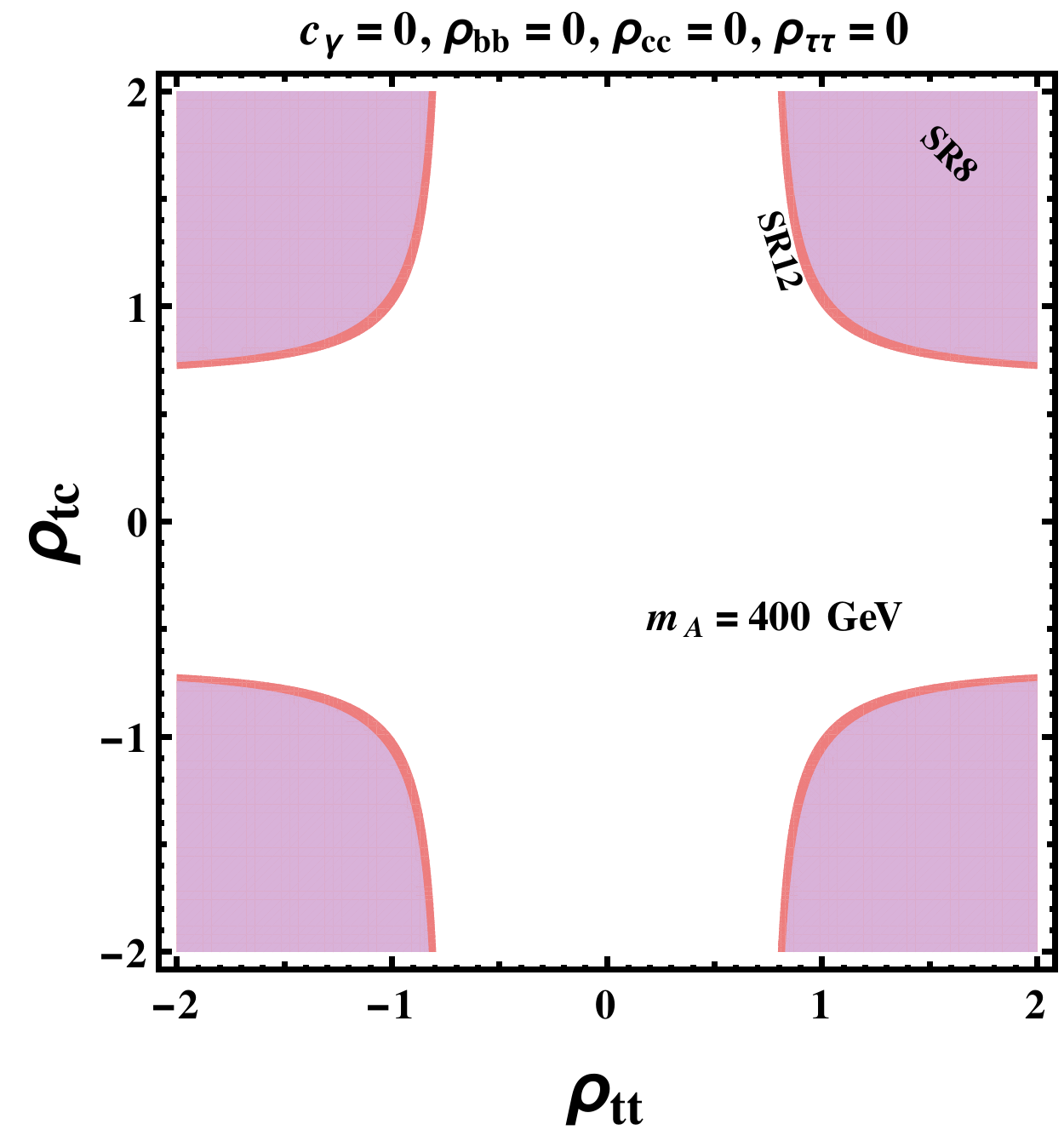}
\includegraphics[width=0.36 \textwidth]{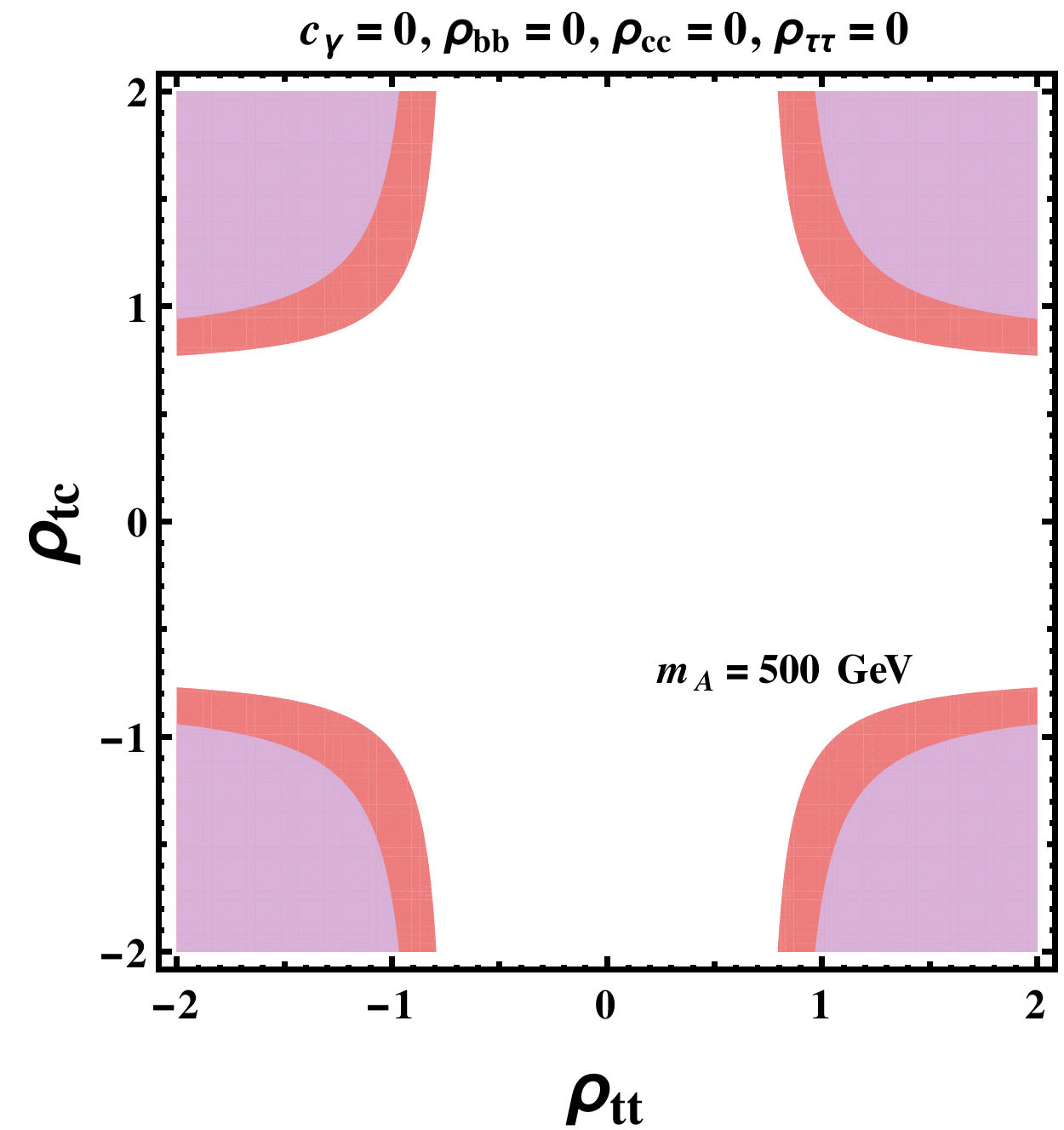}
\caption{
Constraints on $\rho_{tc}$ and $\rho_{tt}$ for $A$ alone case~\cite{Hou:2019gpn}, 
from
 SR8 (purple/light) of Ref.~\cite{Sirunyan:2017roi}, and
 SR12 (red/dark) of Ref.~\cite{Sirunyan:2019wxt}, 
for [left] $m_A = 400$ GeV and [right] 500 GeV.
}
\label{const_nondeg}
\end{figure*}

Skipping the details~\cite{Hou:2019gpn}, we plot in Fig.~\ref{const_nondeg}
the constraints from CMS 4-top search, where
 SR8 is from 36 fb$^{-1}$~\cite{Sirunyan:2017roi}, while
 SR12 is from 137 fb$^{-1}$~\cite{Sirunyan:2019wxt},
for two $m_A$ values, treating it as the lighter exotic boson.
One sees that the constraint has barely improved when the
data has increase four-fold.\footnote{
 At CMS PAS level, SR12 in fact gave worse constraint than SR8.
}
This is largely because for SR12 at 137 fb$^{-1}$, 
CMS has fixed to 4-jets, hence tuned towards 4-top measurement
and giving less improvement on constraining triple-top!

An inadvertent encounter further strengthens our point.
To understand the constraint from $gg \to A/H \to t\bar t$,
we noticed~\cite{Hou:2019gpn} the mention of an ``excess'' (CMS wording) -- but
 neither in Abstract, nor Introduction, nor in Summary -- in CMS-PAS-HIG-17-027, 
though ``deviation'' appeared in abstract of arXiv version~\cite{Sirunyan:2019wph}.
To quote roughly:
``a signal-like excess for the pseudoscalar hypotheses
  (largest) at 400 GeV, $\Gamma/\Gamma_{\rm tot} = 4\%$,
 3.5$\sigma$ local (1.9$\sigma$ look-elsewhere)'',
where we do not quote the CMS plots that illustrate the deviation.
This study has traditionally been viewed by theorists as difficult~\cite{Carena:2016npr},
due to the peak-dip nature of interference with SM $t\bar t$ background. 

Several points are worthy of note~\cite{Hou:2019gpn}:
\begin{itemize}
\item The 400 GeV ``signal'' for $A^0 \to t\bar t$ is consistent with
   $\rho_{tt} \lesssim 1.1$, $\rho_{tc} \lesssim 0.9$
   (CMS did not give actual values of coupling modifiers),
   which is  right on the border of the full Run-2 CMS constraint
   from 4-top (Fig.~\ref{const_nondeg});
   and a dedicated 3-top search would provide further information.
   In our study, $H$ and $H^+$ are at 500 and 550 GeV, respectively.
\item It should be emphasized that 1.9$\sigma$ global significance should
  be ``below radar'', but since ATLAS pioneered the $t\bar t$ scalar resonance
  search~\cite{Aaboud:2017hnm} with Run~1 data (although with $m_{t\bar t}$
  starting at 500 GeV), it would be interesting to see an ATLAS update
  with Run~2 data, and needless to say, the full Run~2 result for both
  ATLAS and CMS.
\item The ``excess'' of CMS, though for pseudoscalar $A^0$, could actually
  arise from scalar $H^0$, because with fully imaginary Yukawa coupling $\rho_{tt}$,
  the $H^0$ could mimic the $A^0$ in the triangle top loop of $gg\to H$ production
  (see Ref.~\cite{Hou:2018uvr} for a brief discussion, for $H^0$ admixture in $h^0$).
\end{itemize}

\section{Conclusion}

With one scalar doublet completed, a second doublet seems rather likely.
With {\it No New Physics} seen so far,
 we should check our {\it Presumptions}.
From this angle, we advocate 2HDM without $Z_2$,
which permits extra Yukawa couplings such as $\rho_{tt}$ and $\rho_{tc}$,
and should be checked experimentally.

With $\rho_{tt} \to 0$ and $\rho_{tc}$ driving EWBG,
 we find relatively light $A^0$ below $t\bar t$ threshold could give
 $tt\bar c$ signature of same-sign top, which is not yet ruled out;
 relatively light $H^0$ below $t\bar t$ threshold could give
 $th^0h^0$ or top-assisted di-Higgs signature.
These signatures should be searched for.

Above $t\bar t$ threshold and with finite $\rho_{tt}$,
while experiments are zooming in on 4-top, 
we continue to advocate direct search of 3-top.
Stumbling on the ``excess'' in $A^0 \to t\bar t$ as reported by CMS,
 we note that this could arise from g2HDM, even from $H^0$ origin!

In conclusion, the exotic scalars of an extended Higgs sector
 may be sub-TeV in mass, and
 far richer and more {\it complex} 
than we are used to think.

\end{document}